\documentstyle[aaspp4,epsfig]{article}
\def\simlt{\lower.5ex\hbox{$\; \buildrel < \over \sim \;$}}
\def\simgt{\lower.5ex\hbox{$\; \buildrel > \over \sim \;$}}
\def\simpt{\lower.5ex\hbox{$\; \buildrel \propto \over \sim \;$}}

\def\kms{\mbox{ km s$^{-1}$}}
\def\mpc{\mbox{ Mpc}}

\def\msun{\mbox{ M}_\odot}

\begin{document}

\title{The Importance of Intergalactic Structure to Gravitationally
  Lensed Quasars}

\author{R. Benton Metcalf\footnote{Hubble Fellow}}
\affil{\it Department of Astronomy and Astrophysics, University of California,
Santa Cruz, CA 95064 USA}

\abstract{
Image flux ratio anomalies have been attributed to substructures
within the gravitational lens and to small mass halos
($M \simlt 10^{10}\msun$) in intergalactic space.   In this paper, analytic calculations
are presented that help in the understanding of how intergalactic halos
affect magnification ratios.  
It is found that intergalactic halos can produce
anomalies at a similar level to those that are observed.
Intergalactic halos with
masses $<10^{10}\msun$ are expected to cause relative deflections between
images of order 10~milliarcseconds, which are then magnified by the primary
lens.  They will also cause fluctuations in the surface density on the
several percent level.  The importance of intergalactic halos depends
strongly on the radial profile of the halos and the primordial power
spectrum at small scales.  Strongly lensed quasars provide an opportunity to probe
these properties. 
A strong dependence on the QSO redshift is predicted and can
be used to distinguish between intergalactic structure and
substructure as the cause of magnification anomalies.
This analytic approach also explains why some previous
semi-analytic estimates disagreed with numerical calculations.  
}

\section{Introduction}

Gravitationally lensed quasars have recently been
used to identify substructure in the dark matter halos of lens galaxies
\markcite{1998MNRAS.295..587M,2001ApJ...563....9M,2002ApJ...567L...5M,2002ApJ...580..696M,2002ApJ...565...17C,Dalal2002,astro-ph/0112038,2003ApJ...584..664K,cirpass2237}({Mao} \& {Schneider} 1998; {Metcalf} \& {Madau} 2001; {Metcalf} \& {Zhao} 2002; {Metcalf} 2002; {Chiba} 2002; {Dalal} \& {Kochanek} 2002; {Brada{\v c}} {et~al.} 2002; {Keeton} 2003; {Metcalf} {et~al.} 2004).
The Cold Dark Matter (CDM) model predicts a certain amount of small
mass subhalos  ($\simlt 10^7 \msun$) that do not have visible dwarf
galaxy counterparts.  These subhalos within the gravitational lens
could be causing the observed anomalies in the image flux ratios, but
it is also possible that they are being caused by clumps of matter
somewhere along the line of sight, but not associated with the lens
galaxies.   \markcite{2004ApJ...604L...5M}{Mao} {et~al.} (2004) have argued that
$\Lambda$CDM does not predict enough substructure within galactic
halos to account for the anomalies.  \markcite{AMCO04}Amara {et~al.} (2004) have 
studied the lensing properties of lenses taken out of CDM simulations
and found that they do not contain enough substructure, but this
conclusion may be dependent on resolution.  
Future cosmological simulation with higher resolution might be able to
verify this for certain.

It was found in \markcite{Metcalf04}Metcalf (2004), using numerical simulations, that
intergalactic halos could account for most of the anomalies seen in
the magnification ratios.  This conclusion has been somewhat
controversial largely because it disagrees with the conclusions of
\markcite{2003ApJ...592...24C}{Chen}, {Kravtsov}, \&  {Keeton} (2003) who found that the contribution from
intergalactic halos would be $\simlt 10\%$ of that from substructure
within the primary lens halo, not enough to account for the observed
anomalies by themselves.  \markcite{2003ApJ...592...24C}{Chen} {et~al.} (2003) relied on an
analytic calculation where the deflection caused by the primary lens
is approximated by a first order expansion and then the cross section
as a function of magnification is calculated for a single perturbing
substructure.  In this paper, it is demonstrated why this type of
calculation is inadequate and how the results of the numerical
simulations \markcite{Metcalf04}(Metcalf 2004) can be understood in terms of a simple
analytic model.

This paper also makes clear what properties of small scale structure
most affect the magnification ratios of strongly lensed quasars.
Previously, the perturbing halos were simply modeled as truncated Singular
Isothermal Spheres (SISs), except in \markcite{Metcalf04}Metcalf (2004), while here
more realistic profiles are used and the sensitivity of lensing to
their structure is investigated.  The analytic model developed here demonstrates
how flux ratio anomalies and Spectroscopic
Gravitational Lensing (SGL) (see \markcite{cirpass2237}{Metcalf} {et~al.} (2004)) can be used to
study the small scale, or small mass, end of structure formation.

For the calculations in this paper the standard $\Lambda$CDM cosmological
model is used with $\Omega_m=0.3$, $\Omega_\Lambda=0.7$, $H_o=70
\kms\mpc^{-1}$ and a scale free initial power spectrum.
In the next section the density profiles and number densities of
small mass dark matter halos are discussed.  In
section~\ref{sec:lens-extr-halos} the importance of intergalactic structure
 to the magnifications of QSO images is calculated both using a halo
model, section~\ref{sec:simplified-model}, and directly from the power
spectrum of density fluctuations, section~\ref{sec:power-spectr-appr}.  The
conclusions are discussed in section~\ref{sec:discussion}.

\section{Halo Profiles and Concentrations}
\label{sec:halo-prof-conc}

From cosmological N-body simulations it has been found that the density
profile of a dark matter halo tends to take a universal form given by (at
least approximately) Navarro, Frenk \& White (NFW) --
$\rho(r)=\frac{\delta_c \rho_c r_s}{r (1+r/r_s)^2}$  
where $\rho_c$ is the critical density of the universe and $r_s$ is the
scale size \markcite{1997ApJ...490..493N}({Navarro}, {Frenk}, \&  {White} 1997).
The concentration of the halo is defined as $c\equiv r_{200}/r_s$ 
where $r_{200}$ is the radius within which the average density is 200
times the mean density of the universe, a proxy for the virial radius of
the halo.  The outer boundary of the halo is generally taken to be
$r_{200}$ as will be the case here.  Once $c$ and the mass, $M$, of a halo are
fixed, $\delta_c$ and $r_s$ can be calculated.  If a relationship
between $c$ and $M$ is specified, the profile is a one parameter family.

The concentration increases with decreasing halo
mass in the CDM model of structure formation.  This is a result of
smaller halos generally forming earlier, when the universe was denser.
\markcite{1997ApJ...490..493N}{Navarro} {et~al.} (1997) propose a relation between concentration and
mass by assuming that when a halo forms it contracts by a fixed factor
from the mean density of the universe at that time.  There is some
ambiguity in defining the collapse time and not all halos of a given mass
form at the same time so this proscription is at best approximate.
More recently, fits to N-body simulations give approximately a power-law
dependence of the concentration on subhalo mass, 
\begin{equation}\label{eq:concentration}
c(M,z) = \frac{c_o}{(1+z)}\left(\frac{M}{10^{12}\msun}\right)^{\beta}.  
\end{equation}
Values for $\beta$ range from $-0.1$ to $-0.2$ and $c_o \simeq 14$
\markcite{2001MNRAS.321..559B,astro-ph/0308348}({Bullock} {et~al.} 2001; {Col{\'{\i}}n} {et~al.} 2004).  In \markcite{Metcalf04}Metcalf (2004)
a redshift independent $c$-$M$ relation was used as in 
\markcite{1997ApJ...490..493N}{Navarro} {et~al.} (1997).  N-body simulations do show that
the concentration of pure dark matter halos of a fixed mass should
have a redshift dependence given above.  This decreases their strength
as lenses somewhat, as will be seen in
section~\ref{sec:lens-extr-halos}.  However, there are a number
of factors the complicate the $c$-$M$ relation and would
be expected to increase the effective concentration of a halo.

In addition to dark matter, halos contain baryons that can significantly
change their profiles.  For example, the contraction of the baryons
because of cooling and angular momentum loss will increase their density
at the center of the halo and drag in additional dark matter, making the
profile steeper and more concentrated.
Lensing results indicate that the inner regions of large galaxies have a
profile that approximates a singular isothermal sphere (SIS),
$\rho(r)\propto r^{-2}$ \markcite{2004ApJ...611..739T}({Treu} \& {Koopmans} 2004).  These effects
are expected to be smaller for small mass halos where the
baryons do not cool efficiently and can be expelled by supernova
feedback \markcite{1986ApJ...303...39D}({Dekel} \& {Silk} 1986).  Observed dwarf galaxies are
known to have higher mass to light ratios than $L_*$ galaxies
\markcite{1998ARA&A..36..435M,2002MNRAS.330..792K}({Mateo} 1998; {Kleyna} {et~al.} 2002).   This also indicates
that the profiles of small mass halos are more 
directly related to early structure formation driven by dark matter
and less confused by feedback from star formation.
 With this in mind, it seems reasonable to approximate the profile of
 small mass halos by the NFW profile with a $c$-$M$ relation
approximating~(\ref{eq:concentration}), but, given the uncertainties
and for the purposes of comparison with \markcite{Metcalf04}Metcalf (2004),
a redshift independent $c$-$M$ relation will also be used in calculations.

The $c$-$M$ relation in N-body simulations also has a large scatter
about the median (\markcite{2001MNRAS.321..559B}{Bullock} {et~al.} (2001) find a 1 $\sigma$
scatter of $\Delta(\log c)\simeq 0.2$).  This is likely to increase
the importance of these halos for lensing since their effects
generally increase with concentration more rapidly than linear.  This
is another reason why expression~(\ref{eq:concentration}) is a
conservative approximation for our purposes.

The mass function of intergalactic halos is calculated using
the Sheth-Tormen \markcite{2002MNRAS.329...61S}({Sheth} \& {Tormen} 2002) modification of the
Press-Schechter \markcite{1974ApJ...187..425P}({Press} \& {Schechter} 1974) method.  This is known to
agree with N-body simulations down to $M\simeq 10^5\msun$
\markcite{astro-ph/0308348}({Col{\'{\i}}n} {et~al.} 2004).  This prescription accounts only for
isolated halos.  All halos that have been resolved in N-body
simulations contain substructure within them.  This hierarchy of
subhalos could conceivably increase the importance of 
intergalactic halos to lensing.  This will not be taken into account
in this paper.

\section{Lensing by Intergalactic Halos}
\label{sec:lens-extr-halos}
\begin{figure}[t]
\centering\epsfig{figure=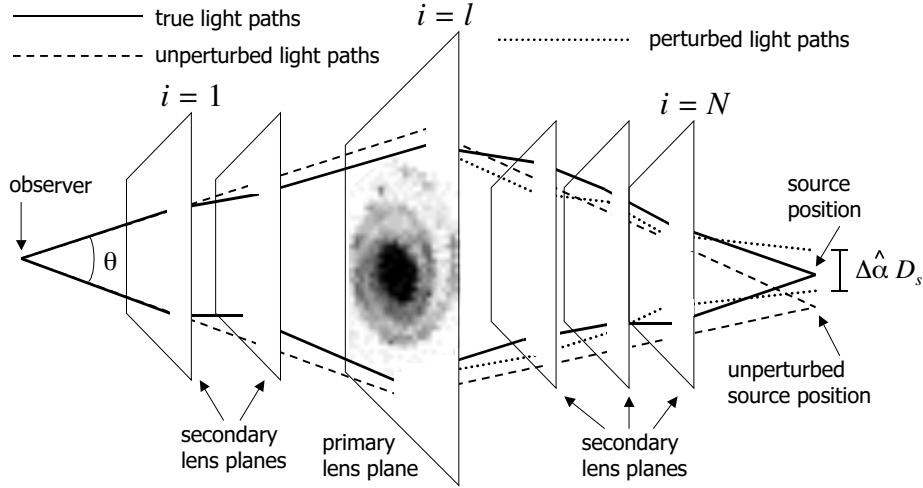,height=5.6in,angle=-90}
\caption[]{\footnotesize A schematic diagram of the type of lensing
  system being considered.  There is one primary lens that is
  responsible for the multiple images of the source.  In addition,
  there are many secondary lenses (most not shown).  The two images
  represented here are separated by an angle $\vec{\theta}$.  There
  may also be two other images that are not shown.  The unperturbed
  light paths are deflected only by the primary lens and with an
  appropriate model for the primary lens will meet on the source
  plane.  If the deflections from secondary lens planes are taken into
account without changing the primary lens model the light will follow
the perturbed light paths (dotted curves) and will not necessarily
meet at a single source.  On the source plane they will be separated
by a distance $\Delta\hat{\alpha}D_s$.  To get the true light paths
the primary lens model needs to be adjusted so that they again meet at
a single point on the source plane.  The angle $\Delta\hat{\alpha}$ is
thus a measure of how wrong the lens model is when intergalactic
lensing is not taken into account.  This diagram is not to scale in
any respect.}
\label{fig:diagram}
\end{figure}

The situation of interest is one where a large primary lens that is creating
multiple images of the QSO and additional small perturbing halos along
the line of sight.  First we will use the multiple lens planes
approximation (see figure~\ref{fig:diagram}).  Here the
deflections caused by each object are
treated as if they take place suddenly in the plane of the lens and
the light follows an unperturbed geodesic between the planes.
This is known to be a very good approximation.  Given the angular
position of a point on the source, $\vec{\beta}$, we must
calculate the image points, $\vec{\theta}$, that correspond to it.
If there are $N$ lens planes these angular positions are related by
\begin{eqnarray}\label{lens_eq}
\vec{\phi}_{j+1}=\frac{D_{j+1}}{D_s}\vec{\theta} - \sum_{i=1}^j
\frac{D_i,_{j+1}}{D_s}\vec{\alpha}_i\left(D_s \vec{\phi_i}\right) ~~~~~~~~
\vec{\beta}= \vec{\phi}_{N+1}(\vec{\theta})
\end{eqnarray}
where $D_i$ is the angular size distance to the $i$th lens, $D_i,_j$ is the
distance between the $i$th and the $j$th lens planes and $D_s=D_{N+1}$ is the
distance to the source.  The position in physical units on the $i$th
lens plane is $\vec{r}_i=D_s \vec{\phi_i}$ and the deflection angle
caused by that lens is $\vec\alpha_i(\vec{r}_i)$.  In general, the
sub- or superscript $l$ will refer to the primary lens.  
The distance to the primary lens is $D_l$ and $D_l,_{N+1}=D_{ls}$.
Equation~(\ref{lens_eq}) is only valid for a geometrically flat
cosmology because it assumes that
$D_i,_{i+1}+D_{i+1},_{i+2}=D_i,_{i+2}$.  We assume that this is the
case in this paper.

The magnification matrix is conveniently decomposed into the
convergence, $\kappa$, and two components of shear, $\gamma_1$ and $\gamma_2$,
\begin{eqnarray}\label{mag_matrix}
A^i_j(\vec{\theta}) \equiv  \frac{\partial\beta^i}{\partial\theta^j} = 
\left(\begin{array}{cc}
1-\kappa-\gamma_1 & \gamma_2 \\
\gamma_2 &  1-\kappa+\gamma_1 
\end{array}\right).
\end{eqnarray}
  When there are multiple lens planes it is possible to get a rotation
, but it will be small and of a
higher order in the perturbations than is considered in this paper.  
In addition, rotation will not affect the magnification ratios.  If there is only
one lens plane, the convergence can be directly related to the local
surface density, $\kappa(\vec{r}_l)=\Sigma(\vec{r}_l)/\Sigma_{\rm
  crit}$.  The critical surface density is $\Sigma_{\rm crit} = (4\pi
G D_lD_{ls}/c^2D_s)^{-1}$.  The magnification of a point-like image is
$\mu=(\det A)^{-1}$.

\subsection{simplified model}
\label{sec:simplified-model}

The influence of intergalactic halos on the magnification ratios and the
cusp caustic relation was studied in \markcite{Metcalf04}Metcalf (2004) by directly solving
equations~(\ref{lens_eq}) in numerical simulations.  Numerical
simulations are necessary for making accurate quantitative predictions, but a
simplified model can help greatly with our understanding of which
parameters and effects are important.

The essential difficulty with solving the lens
equation~(\ref{lens_eq}) is that the positions on each lens plane
are dependent on the positions on the other lens planes.  The strategy
taken here is to calculate quantities
along unperturbed lines of sight, or the path the light would take if
only the primary lens were present.  For an unperturbed path, the
position on the $i$th plane is
\begin{equation}\label{unperturbed}
\vec{r}_i = D_i\vec{\theta} - 
\frac{D_iD_{ls}}{D_s} \left\{
\begin{array}{ccc}
 \vec{\alpha}_l(D_l\vec{\theta}) &~,& i>l \\
0 &~,& i\leq l
\end{array}\right.
\end{equation}
This approximation is valid as long as the deflections by
intergalactic halos are small so that quantities that are second order
in their deflections and derivatives of their deflections can be ignored.

The surface density of the NFW profile can be calculated directly,
but the result is cumbersome. A simplified surface density
can be used that exhibits the essential properties of the NFW profile
and allows for some adjustments
\begin{eqnarray}\label{eq:Sigma}
\Sigma(r,z)=\frac{\Sigma_o}{(1+r/r_s)^n} 
\end{eqnarray}
where $n\simeq2$ is most like the NFW profile although $n=3$ will also
be used to gain some insight into how the profile affects the lensing 
properties.  For $n=2$ the mass within the radius $r$ is
\begin{eqnarray}
M(x)=2\pi r_s^2\Sigma_o \left[
  \ln\left( 1+x \right) - \frac{x}{1+x}
  \right] ~~~~~~ x\equiv \frac{r}{r_s}.
\end{eqnarray}
The total mass is $M_{200} = M(c)$ where the concentration,
$c(M_{200})$, is given by (\ref{eq:concentration}).  For $n=3$ a
different expression for $M(x)$ is derived, but $M_{200}$ and
$c(M_{200})$ are defined in the same way.

Small halos affect the magnification of an image in two ways.  One is
from the halos changing the image positions and the primary lens'
magnification being different at 
the new locations.  Another way is for the substructure to contribute
directly to the magnification of the image by changing
$A^i_j(\vec{\theta})$ without changing the image position enough to
change the host lens's magnification.  One of the goals of this paper
is to determine which of these effects is more important.  For this
purpose the deflection, the convergence and the shear along the
unperturbed path will be considered separately in the next two
subsections.

\subsubsection{deflection}
\label{sec:deflection}

\begin{figure}[t]
\centering\epsfig{figure=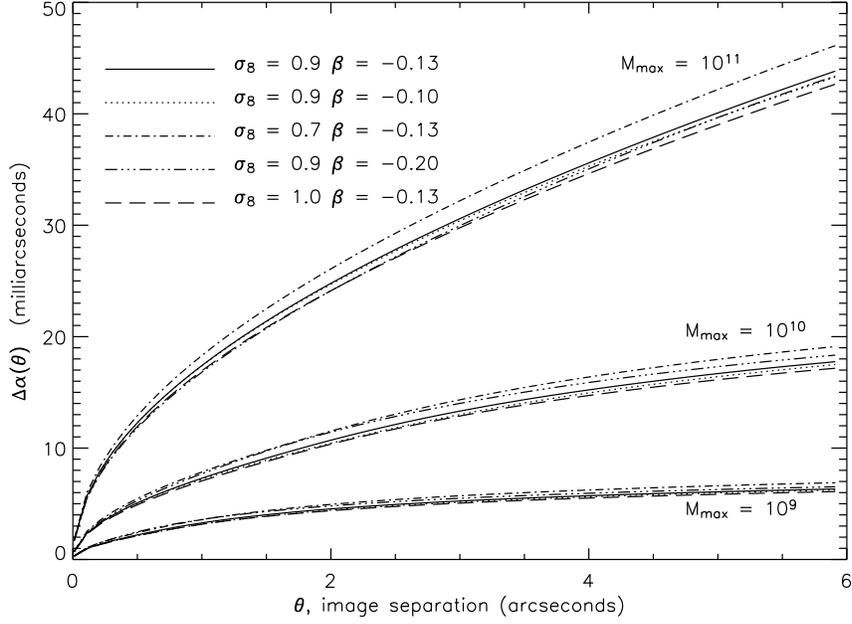,height=3.5in}
\caption[]{\footnotesize The root-mean-squared perturbation to the angular
  separation, $\Delta\hat{\alpha}(\theta)$, as a function of images
  separation, $\theta$.  The range of halo masses included is
  $10^4\msun < M < 10^{11}\msun$ 
for the upper set of curves, $10^4\msun < M < 10^{10}\msun$ for the
middle set  and $10^4\msun < M < 10^9\msun$ for the
lowest set.  In each set there are curves with varying values for the
power spectrum normalization, $\sigma_8$, and the dependence of
concentration on halo mass, $\beta$.}
\label{fig:Dalpha2}
\end{figure}
\begin{figure}[t]

\centering\epsfig{figure=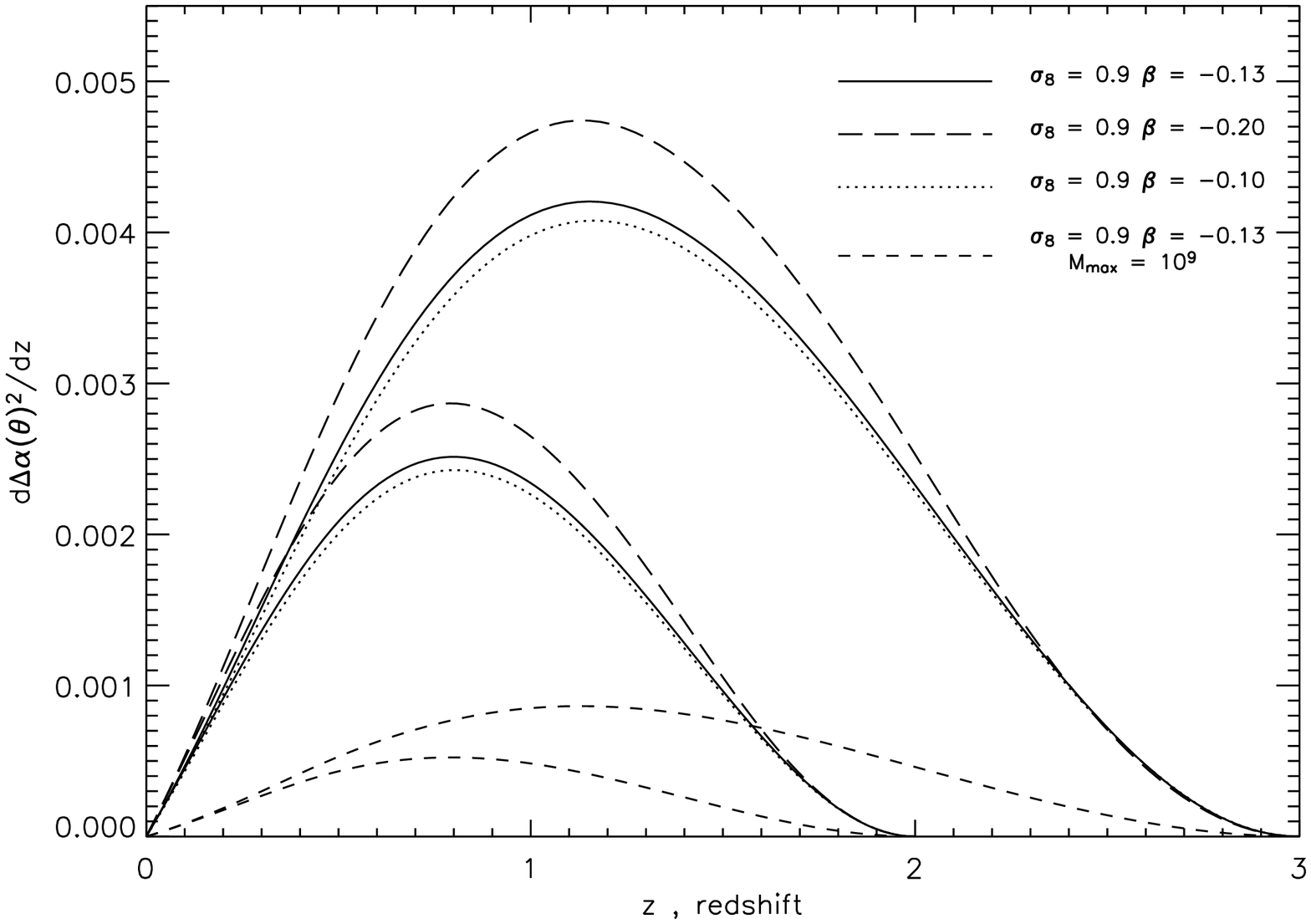,height=3.5in}
\caption[]{\footnotesize The contribution to 
$\Delta\hat{\alpha}(\theta)^2$ from
  intergalactic halos as a function of redshift. The curves that end at
  $z=3$ are for a source at that redshift and a primary lens at
  $z_l=0.5$.  For the others $z_s=2$ and $z_l=0.3$.  The halo mass
  range is $10^4\msun<M<10^{10}\msun$ in all cases except for the
  short dashed curves where $10^4\msun<M<10^9\msun$.  The image
  separation is set to 1~arcsecond.}
\label{fig:dDalpha2dz}
\end{figure}

The deflection angle caused by a axisymmetric halo is 
\begin{eqnarray}\label{eq:deflection}
\vec{\alpha}(\vec{r}) = 4 G M\left(r/r_s\right)\frac{\vec{r}}{r^2}.
\end{eqnarray}
From the lens equation~(\ref{lens_eq}) it is seen that the important
quantity is $\hat{\alpha}(r)=\frac{D(z,z_s)}{D_s}\vec{\alpha}(r)$
which is the lowest order displacement of a light ray on the source
plane (see figure~\ref{fig:diagram}).

Since $|\alpha(r)|$ falls only as $r^{-1}$, for large $r$ it might be
expected that the deflections of many
halos will add up to a total that is substantially larger than what an
individual halo is likely to cause by itself.  In fact, 
$\langle\hat{\alpha}(\theta)^2\rangle$ diverges logarithmically in the
plane lens approximation because the number of halos rises as $r^2$.  
This is not a problem here however because only differences in the
deflection angle are important.
A constant deflection over the entire lens is equivalent to a shift
in the position of the source.  Such a uniform shift would be taken
into account when fitting a primary lens model so it cannot produce
any anomaly in the magnification ratios.  However, if the deflections
are different for the different images then they will cause the
image positions to shift independently.

To asses the significance of these deflections we calculate the
correlation function between the perturbations to the deflection caused
by intergalactic halos as a function of the 
image separation.  The positions of the halos will be treated as random
and uncorrelated.  In this case the correlation is
\begin{eqnarray}
\Delta\hat{\alpha}(\theta)^2 & \equiv & \left\langle
\left[ \hat{\alpha}(\vec{\theta}')  - \hat{\alpha}(\vec{\theta}' +
  \vec{\theta}) \right]^2 \right\rangle \label{dalpha1} \\ \label{dalpha2}
&=& \frac{1}{H_oD_s^2} \int_0^{z_s}
dz~\frac{D(z,z_s)^2}{\sqrt{\Omega(1+z)^3+\Omega_\Lambda}}
\int_0^{M_{\rm max}} dM~
\frac{dN}{dM}(M,z) \\ \nonumber
& & ~~ \times \int d^2r~|\vec{\alpha}(\vec{r}) - \vec{\alpha}(\vec{r}+\vec{y}(z,\theta))|^2
\end{eqnarray}
where $\vec{y}(z,\theta)$ is the physical separation of two unperturbed
lines of sight given by (\ref{unperturbed}).  These integrals are
calculated numerically.  The range of the $r$ integral in
(\ref{dalpha2}) is increased until the answer stops changing. 

Figure~\ref{fig:Dalpha2} shows $\Delta\hat{\alpha}(\theta)$ for a source
at $z_s=3$ and lens at $z_l=0.5$.  The most important parameter for
$\Delta\hat{\alpha}(\theta)$ is the upper halo mass cutoff, $M_{\rm max}$.  The
concentration and $\sigma_8$ are relatively unimportant.  Using a 
redshift independent $c$-$M$ relation gives slightly larger values for
$\Delta\hat{\alpha}(\theta)$, but it is much smaller than the
dependence on $M_{\rm max}$.  Figure~\ref{fig:dDalpha2dz} what
redshifts are contributing to $\Delta\hat{\alpha}(\theta)$.  Four a
QSO at $z=3$ the majority of the deflection is coming from halos
above $z=1$.

The deflections shown in figure~\ref{fig:Dalpha2} can have a
significant effect on the image magnifications even if the
convergence and shear caused by the substructures are small.  This is
because, for a single image, the perturbation to the deflection acts
like a {\it local} change in the source position.  
The change in the local source position can be magnified by the host lens
to resulting in a substantial change in the image position.
Even if the intergalactic halos did not affect the local magnification
as a function of image position they would affect where the images form
and thus change the magnification caused by the primary lens.  This is
an effect that is 
completely absent from a cross section type calculation where the host
lens' deflection is expanded to first order (constant magnification).
Local changes in the source position on the order of 15 milliarcseconds do
cause changes in the magnification ratios even in Einstein
cross type lenses.  However, as will be shown in
section~\ref{sec:convergence--shear}, the intergalactic halos will
modify the magnification directly as well and probably this second
effect is more important.

The clustering of the halos has not been taken into account in this
calculation.  Clustering into filaments and other structures will only
increase the value of $\Delta\hat{\alpha}(\theta)^2$.  Because the
deflection caused by a halo has such a long range the increase
in $\Delta\hat{\alpha}(\theta)^2$ could be significant, but it will
not be investigated here.

\subsubsection{convergence \& shear}
\label{sec:convergence--shear}
\begin{figure}[t]
\vspace{-1,5cm}
\centering\epsfig{figure=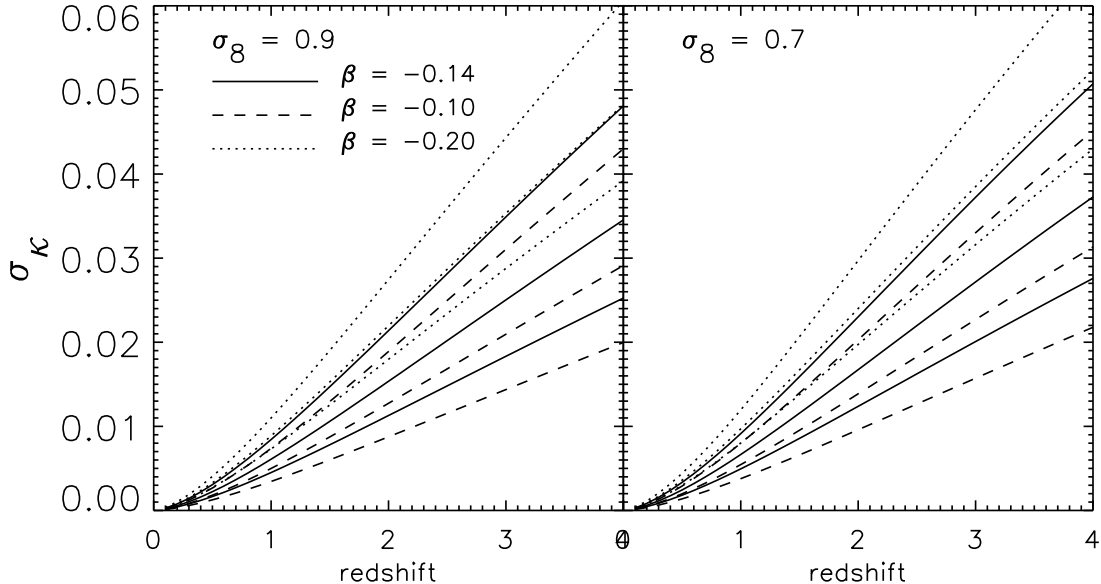,height=4.75in}
\vspace{-1,5cm}
\caption[]{\footnotesize 
The standard deviation in the dimensionless surface density in small scale
structure along random lines of sight as a function of source redshift.  
For the right panel the power spectrum normalization is $\sigma_8=0.9$
as measured by \markcite{2003ApJS..148..175S}({Spergel} {et~al.} 2003) and on the left a lower
normalization is used $\sigma_8=0.7$.  The halo mass-concentration
relation is specified by the $\beta$ values for each type of curve.
For each set of like curves the upper one is for a halo mass range of
$10^3\msun<M<10^{11}\msun$, the middle one is for
$10^3\msun<M<10^{10}\msun$ and the lower one is for $10^3\msun<M<10^{9}\msun$.
}
\label{fig:kappa2}
\end{figure}

\begin{figure}[t]
\vspace{-1,5cm}
\centering\epsfig{figure=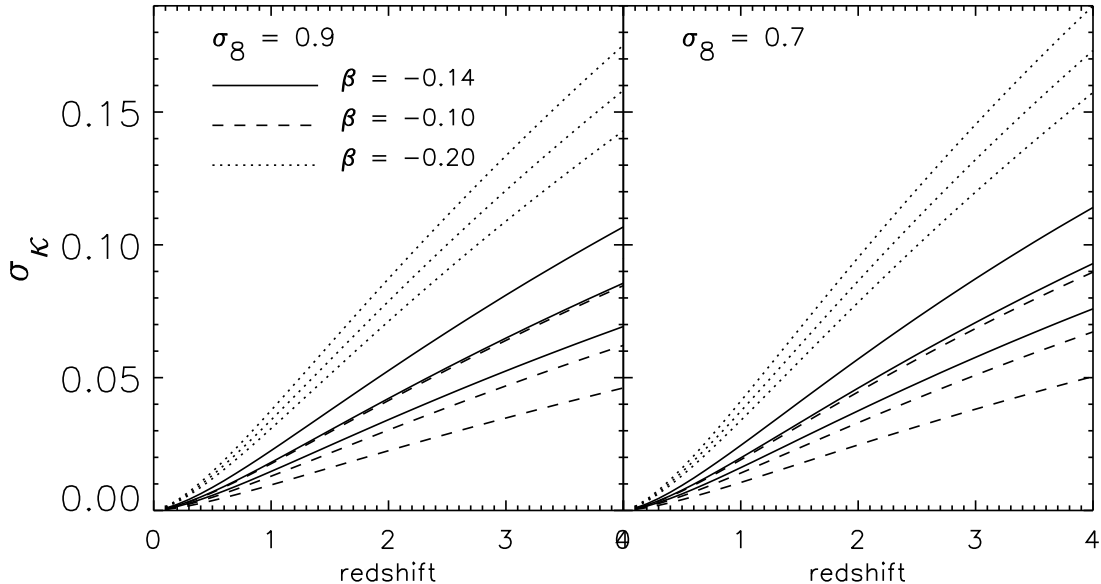,height=4.75in}
\vspace{-1,5cm}
\caption[]{\footnotesize 
The standard deviation in the dimensionless surface density in small scale
structure.  Everything is the same as in figure~\ref{fig:kappa2} except
that the halo profile is modified so that it is cutoff more sharply,
$n=3$ in equation~(\ref{eq:Sigma}).  The variance is significantly larger.
}
\label{fig:kappa2n3}
\end{figure}

\begin{figure}[t]
\vspace{-1,5cm}
\centering\epsfig{figure=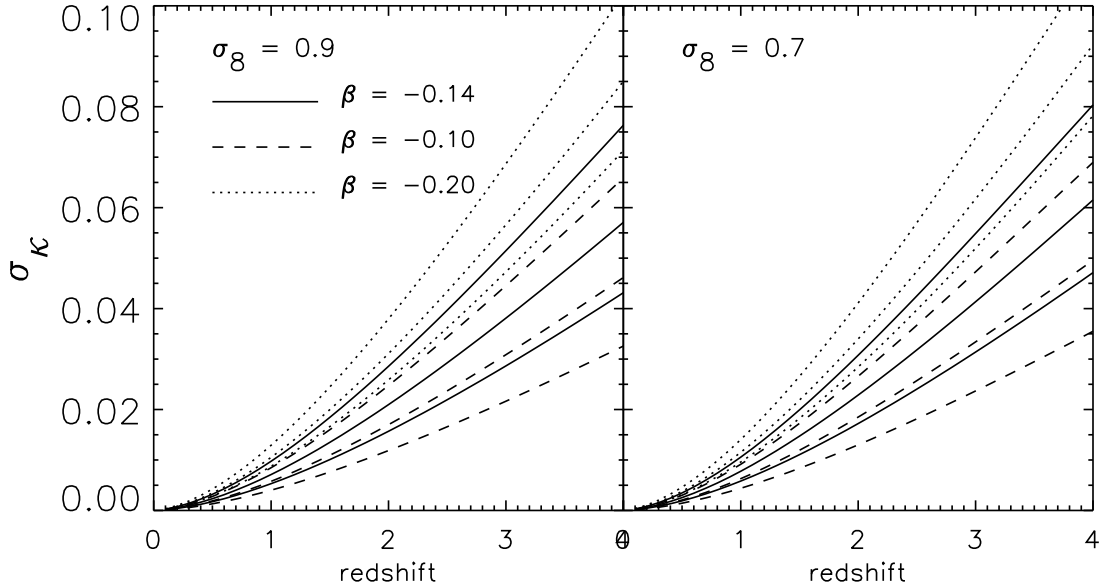,height=4.75in}
\vspace{-1,5cm}
\caption[]{\footnotesize 
The standard deviation in the dimensionless surface density in small scale
structure.  Everything is the same as in figure~\ref{fig:kappa2} except
that the mass--concentration relation is given by
equation~(\ref{eq:concentration}) without the redshift dependence and
with $c_o=12$.  The redshift dependence tends to reduce
$\sigma_\kappa$ somewhat.
}
\label{fig:kappa2_noZ}
\end{figure}

\begin{figure}[t]
\centering\epsfig{figure=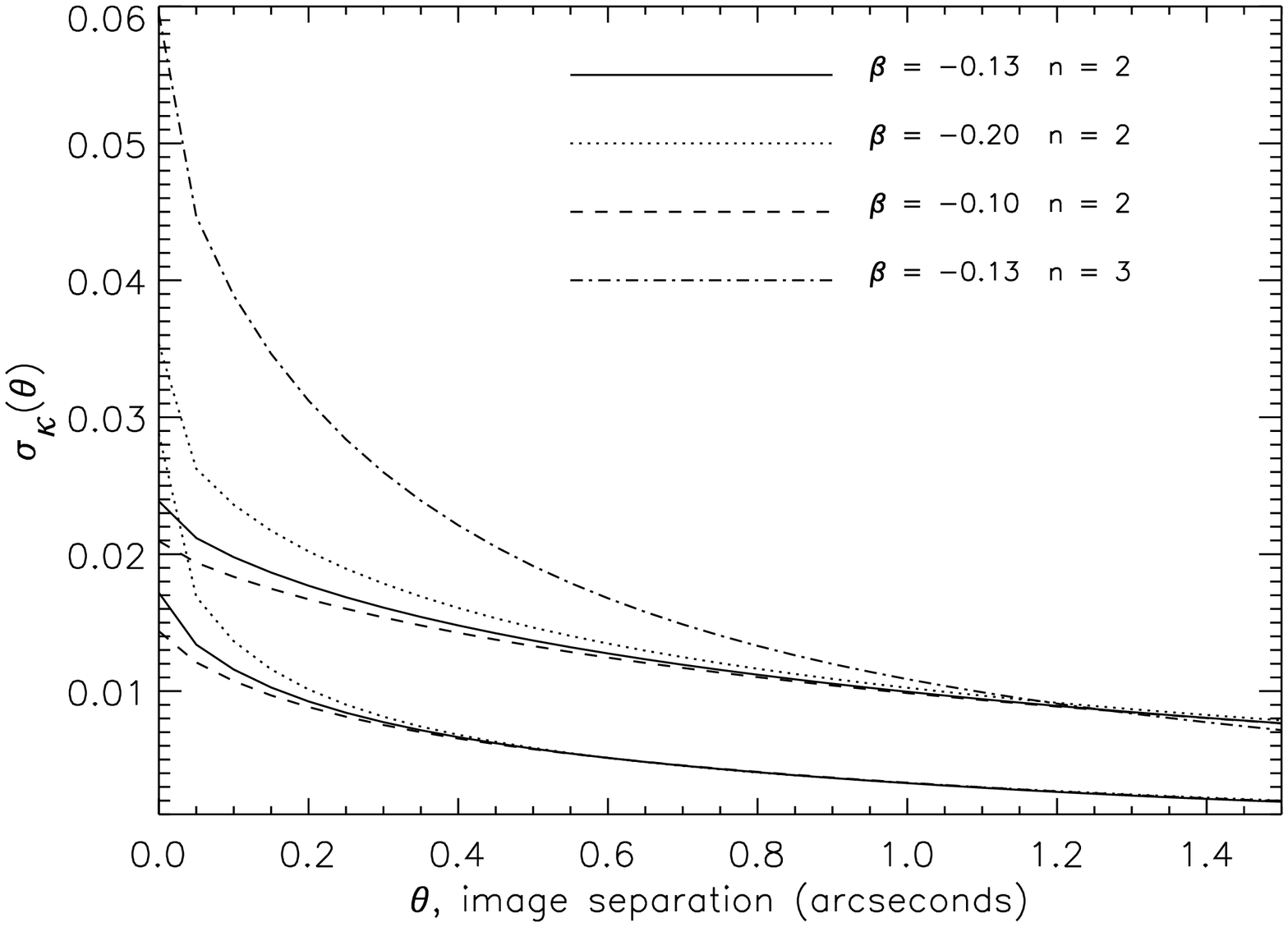,height=3.5in}
\caption[]{\footnotesize The correlation of surface density as a
  function of angular separation between images.  The values for the halo parameters  
  $\beta$ and $n$ are listed.  The three lowest curves (for most of the
  $\theta$-range) are for halo mass ranges $10^3\msun<M<10^9\msun$ and
  the others are for $10^3\msun<M<10^{10}\msun$. The source is taken
  to be at $z_s=3$ and the primary lens is at $z_l=0.5$.  The
  power spectrum normalization is $\sigma_8=0.9$ in all cases.  Only
  one $n=3$ curve is included for clarity of the plot.}
\label{fig:sigk_theta}
\end{figure}

By taking an unperturbed light path in the lens equation~(\ref{lens_eq})
and the then calculating its derivatives it is easily shown that, to
lowest order, the total convergence, $\kappa$, is the sum of all the
surface densities of all the halos along the line of sight weighted by
the critical density, 
\begin{equation}
\kappa(\theta) \simeq 
\sum_i \frac{\Sigma_i(\vec{\theta})}{\Sigma_{\rm crit}(z_i,z_s)} 
= \sum_i \kappa_i(\vec{\theta}).
\end{equation}
In the standard halo model of structure in the universe, all mass is
in halos of some size.  The standard expression for the angular size
distance in a homogeneous universe takes into account the average
surface density so, what is important for perturbations in the image
magnifications is the variation in $\kappa(\vec{\theta})$ about the
mean.  In addition, if the perturbation is uniform over the entire lens
area, it will not affect any observables because of the mass sheet
degeneracy.  To cause anomalies in the magnification ratios
$\kappa(\theta)$ must have significant fluctuations on a scale that is
smaller than roughly the Einstein ring radius of the primary lens
($\sim 1$ arcsecond).

From the halo profile and the mass function the variance in 
$\kappa$ along an unperturbed line of sight, $\sigma^2_\kappa \equiv
\langle \kappa^2\rangle - \langle \kappa\rangle^2$, can be
calculated assuming uncorrelated halo positions in the same manner as
the deflection angle was calculated in section~\ref{sec:deflection},  
\begin{eqnarray}\label{eq:kappa2}
\sigma_\kappa^2(\theta) 
&=& \frac{1}{H_o} \int_0^{z_s}
dz~\frac{\Sigma_{\rm crit}(z,z_s)^{-2}}{\sqrt{\Omega(1+z)^3+\Omega_\Lambda}}
\int_0^{M_{\rm max}} dM~ \frac{dN}{dM}(M,z) \\ \nonumber
& & ~~ \times \int_0^{\infty} d^2r~\Sigma\left(\vec{r},M,z\right)\Sigma\left(\vec{r}+\vec{y}(z,\theta),M,z\right)
\end{eqnarray}
The mass integral is bounded on the top by the minimum mass of a halo that one
would expect to have an observable galaxy inside it.  Any observed
galaxies would be included in the primary lens model.

Figure~\ref{fig:kappa2} shows the result for $\sigma_\kappa(0)$ as a
function of source redshift.   It can immediately be seen that the
contribution of intergalactic halos 
to millilensing can be significant.  Previous estimates for the amount
of substructure required to account for the monochromatic magnification
anomalies were a few percent.  Given that the surface density in the
primary lens is $\kappa\simeq 0.5$ at the critical curve near where the
images are located, the surface densities in figure~\ref{fig:kappa2} are
substantial.  Just like for $\Delta\hat{\alpha}(\theta)$,
$\sigma_\kappa(\theta)$ will only increase when the clustering of the halos is
included.

By comparing figures~\ref{fig:kappa2} and \ref{fig:Dalpha2}, it is seen
that $\sigma_\kappa(\theta)$ has a stronger dependence on smaller mass
halos, while $\Delta\hat{\alpha}(\theta)$ is heavily weighted toward
the high end of the mass range.
One can also see in figure~\ref{fig:kappa2} that the concentration of
these halos has an important influence on their lensing properties. 
The redshift dependence is also important.  A way to 
distinguish intergalactic halos from substructure in the primary lens is
to measure how the anomalies depend on source redshift.  Since the
space density of small scale halos becomes larger at higher redshift there is
a significant increase in $\sigma_\kappa(z)$ over the range of
observed QSOs lenses ($z\sim 1$ to 4).
The dependence of the normalization of the matter power spectrum may be
counter intuitive to some.  If $\sigma_8$ is reduced, more of the mass is
in smaller halos and since we are only including small halos (well below
$M_*$, the nonlinear mass scale), $\sigma_\kappa$ is increased.  
If the current measurements of $\sigma_8$ are taken into
consideration and the primordial power spectrum is assumed to be scale
invariant down to these scales, this
effect is relatively small compared with the importance of halo
concentration.  
If scale invariance is broken however this could become an important dependency.

A very interesting and illuminating thing occurs when $n\neq 2$.
Figure~\ref{fig:kappa2n3} shows the same quantities as
in figure~\ref{fig:kappa2} only with the large $r$ slope changed to
$n=3$.  This has the effect of cutting
the halos off more abruptly and, to conserve mass, increasing their
central densities.  It is clearly seen that $\sigma_\kappa$ is larger in
this case.  This strong dependence on $n$ demonstrates that it is not
the inner core of a single halo that is doing most of the lensing, as is implicitly
assumed in a cross section type calculation.  The important property is
the profile at radii where the halos are likely to overlap on the sky.
A cross section type calculation does not account for this since there
the line of sight is assumed to intersect with at most one halo.

Figure~\ref{fig:kappa2_noZ} shows $\sigma_\kappa(z)$ with a redshift
independent $c$-$M$ relation.  The same $c$-$M$ relation was used in
the simulations of \markcite{Metcalf04}Metcalf (2004).  The fluctuations in $\kappa$
are similar, but larger in this case.  It is possible that baryonic
contraction in the larger halos could convert the inner region of
their profiles into the SIS form ($\rho(r)\propto r^{-2}$) at some
high redshift.  If this is the case, $\sigma_\kappa(z)$ will be even
larger.  If the $z=0$
normalization can be established by other means, the redshift
dependence of the magnification ratio anomalies can be used to
measure the redshift dependence of the $c$-$M$ relation.

As mentioned before, if the fluctuations in $\kappa(\theta)$ have a
coherence length that is much larger than the size of the primary
lens, they will have no observable effects because of the mass
sheet degeneracy.  Figure~\ref{fig:sigk_theta} shows
$\sigma_\kappa(\theta)$ as a function of image separation.  Most of
the fluctuations caused by halos of mass $<10^{10}\msun$ are on a
significantly smaller scale than typical image separations.  The
concentrations of the halos appear to strongly affect only the small scale
fluctuations below a few tenths of an arcsecond. 

One might be concerned that the variance in $\kappa(\vec{\theta})$ is
not a good quantity for evaluating variations.  Often, 
distributions in lensing have long tails which make the variance misleadingly
large -- for example the distribution of microlensing magnification
caused by point masses.  This is not the case here because the halo
profile used does not diverge at the center or fall off too slowly at
large $r$.  This makes $\sigma_\kappa$ a trustworthy indicator of
typical variations.  It is possible that the halos are highly
clustered which would make $\sigma_\kappa$ larger and skew the
distribution so that the mode lies below the mean.  If this were a
dramatic effect the standard expression for the angular size distance
would not apply to most high redshift objects.


The variance in the shear along random lines of sight, defined as
$\sigma^2_\gamma= \langle \gamma_1^2+\gamma_2^2 \rangle$
, can be calculated in the same way as
$\sigma_\kappa^2$ was calculated.  However, the requirement that the
distribution of halo positions be isotropic automatically guarantees
that $\sigma^2_\gamma(\theta) =  \sigma^2_\kappa(\theta)$.  This is best demonstrated in
Fourier space so the explanation will be deferred until
section~\ref{sec:power-spectr-appr}.  Despite this, an explicit
calculation of $\sigma^2_\gamma(0)$ has been done with the halo model and it is
found to equal $\sigma^2_\kappa(0)$ to very high accuracy which provides a
reassuring consistency check.

The perturbations to the magnification matrix represented in
figure~\ref{fig:kappa2} are on the several percent level.  It might seem
like this is too small to be of importance, but in fact these perturbations can
strongly affect the magnifications of the images.  To see this, the
magnification can be expanded around the value for the primary lens alone, $\mu_l$,\footnote{This expansion is not guaranteed to converge when the source
  is very near to a
caustic, but if, roughly speaking, $\mu_l\delta\kappa < 1$ this should not
be a problem.}
\begin{equation}
\frac{\Delta\mu}{\mu_l} \simeq 2\mu_l\left[(1-\kappa_l)\delta\kappa + \vec{\gamma}_l\cdot\delta\vec{\gamma}\right].
\end{equation}
The variance in the perturbation to the magnification is
\begin{equation}
\left( \frac{\Delta\mu}{\mu_l} \right)_{\rm rms} \simeq
2\mu_l\sigma_\kappa(0) \sqrt{(1-\kappa_l)^2 + \frac{\gamma_l^2}{2}}
\end{equation}
where the fact that $\sigma^2_\gamma(\theta) =
\sigma^2_\kappa(\theta)$ has been used. 
Galactic sized lenses are well fit by singular isothermal ellipsoid
(SIE) models, in which case the images form at $\kappa_l \approx \gamma_l
\approx 0.5$.  In this case $\left( \frac{\Delta\mu}{\mu_l} \right)_{\rm
  rms} \approx \mu_l\sigma_\kappa$.  For an image with a modest magnification
of 5 this will exceed 10\% if the source is at $z_s \simgt 2$.  For more
highly magnified images and/or higher redshift sources this can be
significantly larger.  And if halos have a steeper cutoff than NFW at
large $r$ it can be much larger.  The level of the observed anomalies
is on the $\sim 10\%$ level \markcite{2002ApJ...567L...5M}({Metcalf} \& {Zhao} 2002). And direct numerical
simulations give similar results \markcite{Metcalf04}(Metcalf 2004).

\subsection{power spectrum approach}
\label{sec:power-spectr-appr}

The second order statistics calculated thus far using the halo model
can also be calculated in terms of the power spectrum of density
fluctuations.  This may be a more 
direct, although somewhat more obscure, way of interpreting the
observations since the halo model contains more information than do
$\sigma_\kappa^2(\theta)$ or $\Delta\hat{\alpha}(\theta)^2$.  The
power spectrum also has the advantage of including within it the
correlations in halo positions which were ignored in the previous
section and it can be calculated relatively easily from a N-body
simulations .  With enough data one could think of inverting the
variance in magnification anomalies to get a constraint on the power
spectrum at small scales.

The deflection along a short section of the light path is
\begin{equation}
\vec{\alpha}(\vec{\theta}) = 2 \vec{\nabla}_\perp
\Phi\left(\vec{x}(\vec{\theta},z)\right) \delta l
\end{equation}
where $\Phi(\vec{x})$ is the Newtonian potential,
$\vec{\nabla}_\perp$ is the gradient operator in the two dimensions
perpendicular to the light path and $\delta l$ is the path length.  
If the unperturbed path is substituted in for $\vec{x}(\vec{\theta},z)$, the
potential is Fourier transformed, and all the deflections along the
path are added up one gets
\begin{equation}\label{eq:alpha_power}
\hat{\alpha}(\vec{\theta}) = (8\pi G\rho_o) \int dR~
(1+z)\frac{D(z,z_s)}{D_s}\int
\frac{d^3k}{(2\pi)^3}~\frac{\vec{k}_\perp}{k^2} \delta(k)
\exp\left(-i\vec{k}_\perp\cdot\vec{x}_\perp(\vec{\theta}) - i k_rR\right)
\end{equation}
where $\delta(k)$ is the Fourier transform of
$(\rho-\overline{\rho})/\overline{\rho}$, $\rho_o$ is the current
density of the universe and $R$ is the radial distance.  The correlation
function for the deflection and any of its derivatives can be calculated
with the Fourier version of Limber's equation \markcite{Kais92}({Kaiser} 1992) as is commonly
done in weak lensing calculations.  This is essentially the assumption
that the quantities $z$ and $D(z)$ change less rapidly with $R$ than
$\delta(\vec{x})$ does.  The result is
\begin{equation}
\Delta\hat{\alpha}(\theta)^2 =
\frac{4\pi}{H_o}\left(\frac{4G\rho_o}{D_s}\right)^2
\int_0^{z_s} dz~\frac{(1+z)^2D(z,z_s)^2}{\sqrt{\Omega_m(1+z)^3+\Omega_\Lambda}}
\int_0^\infty \frac{dk}{k}\left( 1-J_o[ky(z,\theta)] \right) P(k,z,M_{\rm max})
\end{equation}
where $P(k,z)$ is the power spectrum of density fluctuations and $J_o[x]$
is the zero-th order Bessel function.  And by taking derivatives of
(\ref{eq:alpha_power}), squaring and averaging, one can work out the
variance of all the parts of the magnification matrix.  Specifically
\begin{equation}
\sigma^2_\kappa(\theta) =\sigma^2_\gamma(\theta) = \frac{\rho_o^2}{2\pi H_o}\int_0^{z_s}
dz~\frac{(1+z)^2\Sigma_{\rm
    crit}(z,z_s)^{-2}}{\sqrt{\Omega_m(1+z)^3+\Omega_\Lambda}} 
\int_0^\infty dk~ k J_o[ky(z,\theta)]P(k,z,M_{\rm max}).
\end{equation}
The first equality comes from the fact that
$(k_1^2-k_2^2)^2/4+k_1^2k_2^2=(k_1^2+k_2^2)^2/4=k^4/4$. 

The power spectrum used here should really be a conditional power
spectrum.  As seen in equation~(\ref{eq:deflection}), the deflection
angle is proportional to the mass of the halo so the dispersion can be
dominated by high mass halos. If no other galaxies are seen near the
primary lens or they are included in the lens model for the primary lens
then there is an upper halo mass (or light) cutoff, $M_{\rm max}$,
implicit in this power spectrum.  In practice, $M_{\rm max}$ should be
a function of $z$ to express the detection limits of the observations.

\section{Discussion}
\label{sec:discussion}

It was demonstrated here that intergalactic matter is expected to
cause changes in the magnification ratios that are of the same size
as those that have been observed.  This confirms, at least
qualitatively, the conclusions derived from numerical simulations
\markcite{Metcalf04}(Metcalf 2004).  The possibility that the anomalies are from
substructures inside the primary lens may not be consistent with
$\Lambda$CDM \markcite{2004ApJ...604L...5M,AMCO04}({Mao} {et~al.} 2004; Amara {et~al.} 2004), although simulations with
high enough resolution to conclusively rule this out have not yet been
done.  Probably the best hope for distinguishing observationally between
these two possibilities is to measure the dependence of
magnification ratio anomalies on source redshift in a statistical way.

It was found here that the importance of intergalactic halos to the
magnification ratios depends on the internal structure of the halos
as well as the primordial power spectrum.  The mass--concentration
relation is of particular importance.  This relation depends on the
formation time of the halos (halos that form earlier are denser), the
history of mass accretion and, in turn, on the slope of the
primordial power spectrum and amount of dark energy.  The
concentrations could also change if the dark matter particle has a
small mass (warm or tepid dark matter) or if it is self-interacting.

There are several factors that have not been taken into account here
and would be expected to increase the importance of intergalactic
structure.  The halos will be strongly clustered which will clearly
increase $\sigma_\kappa$.  The scatter in the $c$-$M$ would be
expected to increase halos' influence on lensing as a population.  The
contraction of baryons within the halos will increase
their concentration.  Observed lenses (at a higher mass scale) have
profiles closer to a SIS profile than the NFW profile.  SIS halos
will produce larger discrepancies in the magnifications.  Finally,
intergalactic halos will have substructure 
of their own within them which will increase $\sigma_\kappa$.  For
these reasons the numbers given in this paper are conservative estimates.

To go beyond the variance in the image magnifications and
to go beyond the unperturbed light path approximation numerical
simulations are required.  It has also been assumed here that the
source is infinitely small which in many cases will not be a good
approximation.  One expects that when the size of the source is on the
order of the size of the halos causing the lensing, the variations in
the magnification will be smaller.  Further simulations of specific
lenses are needed.

Instead of thinking of this effect as lensing by individual galaxies in
    intergalactic space it might be
    better to think of it as a variation in the angular size distance as
    a function of position on the sky or a variation in the surface
    density of the universe.  A typical line of sight travels
    through many halos (500 with $10^7\msun < M < 10^9\msun$ for
    $z_s=3$!).  On average they add up to the average 
    surface density of the universe and the normal expression for the
    angular size 
    distance calculated with the Robertson--Walker metric is
    recovered.  However, there are variations in the surface 
    density at the several percent level and these are enough to change
    the magnification ratios of multiply imaged quasars.
This opens up an opportunity to probe the small scale, otherwise
unobservable, structure in the universe which is shaped by the physics
of the early universe as well as the history of star
formation and reionization.

\section*{Acknowledgments}
\footnotesize
I would like to thank J. Primack and R. Somerville for insight into the $c$-$M$
relation and M. Magliocchetti for helpful comments.
Financial support was provided by NASA through Hubble
Fellowship grant HF-01154.01-A awarded by the Space Telescope Science
Institute, which is operated by the Association of Universities for Research 
in Astronomy, Inc., for NASA, under contract NAS 5-26555

\end{document}